\documentclass[reprint,aps,prl,amsmath,longbibliography,superscriptaddress]{revtex4-1}
\usepackage{color}
\usepackage{graphicx}
\usepackage{dcolumn}
\usepackage{bm}
\usepackage{float}
\usepackage{siunitx}
\usepackage{braket}
\usepackage{ mathrsfs }
\usepackage[dvipsnames]{xcolor}
\usepackage{bbold}

\newcommand{\ttr}{\mathrm{Tr}}
\newcommand{\cA}{\mathcal{A}}
\newcommand{\cN}{\mathcal{N}}

\newcommand{\AN}{\mathcal{AN}}

\usepackage[colorlinks=true,
  urlcolor=blue,
  linkcolor=blue,
  citecolor=blue]{hyperref}

\begin{document}
\title{Measurement of Identical Particle Entanglement and the Influence of Antisymmetrisation}
\author{J. H. Becher}
\affiliation{Physics Institute, Heidelberg University,\\
Im Neuenheimer Feld 226, 69120 Heidelberg, Germany}
\author{E. Sindici}
\affiliation{Department of Physics and SUPA, University of Strathclyde,\\
Glasgow G4 0NG, UK}
\author{R. Klemt}
\affiliation{Physics Institute, Heidelberg University,\\
Im Neuenheimer Feld 226, 69120 Heidelberg, Germany}
\author{S. Jochim}
\affiliation{Physics Institute, Heidelberg University,\\
Im Neuenheimer Feld 226, 69120 Heidelberg, Germany}
\author{A. J. Daley}
\affiliation{Department of Physics and SUPA, University of Strathclyde,\\
Glasgow G4 0NG, UK}
\author{P. M. Preiss}
\affiliation{Physics Institute, Heidelberg University,\\
Im Neuenheimer Feld 226, 69120 Heidelberg, Germany}

\date{\today}
      
\begin{abstract}

We explore the relationship between symmetrisation and entanglement through measurements on few-particle systems in a multi-well potential. In particular, considering two or three trapped atoms, we measure and distinguish correlations arising from two different physical origins: antisymmetrisation of the fermionic wavefunction and interaction between particles. We quantify this through the entanglement negativity of states, and the introduction of an antisymmetric negativity, which allows us to understand the role that symmetrisation plays in the measured entanglement properties. We apply this concept both to pure theoretical states and to experimentally reconstructed density matrices of two or three mobile particles in an array of optical tweezers.
\end{abstract}

\maketitle

Entanglement is one of the foundational properties of quantum mechanical systems. On a fundamental level, it helps to capture what is unusual about measurement on quantum mechanical systems, and provides the possibility to violate local realism through measurements of Bell's inequalities \cite{RevModPhys.86.419}. On a more practical level, it is seen as a resource, including for quantum computation, quantum enhanced metrology, and a range of other technologies \cite{Strobel424,RevModPhys.90.035005}. When we discuss entanglement of many-particle systems, there is a range of possible choices for how entanglement may be characterised, and we need to ask for which degrees of freedom we are interested in entanglement properties.
A possible choice is to consider mode entanglement, dividing a system into spatial (or momentum) modes \cite{Tichy_2011}. This is useful in characterising many-body quantum systems, including identifying and studying phase transitions \cite{Amico2008, Eisert2010}, determining classes of many-body quantum states that are simulable on a classical computer \cite{PhysRevLett.91.147902,PhysRevLett.100.030504,Schuch_2008}, and identifying topological states \cite{Kitaev2006, Hui2008, Zeng2019}. However, this is clearly not the only way to define entanglement, and has a pathological limit in terms of extracting useful entanglement, as a single particle moving in space provides entanglement between different spatial modes. \\
An alternative possibility is to consider entanglement between particles. This presents a particular challenge in the case of identical particles, because of their exchange symmetries. For example, given that all electrons in a single atom are identical fermions, we can ask whether their antisymmetry must necessarily imply that they are all entangled with each other \cite{Schliemann01, Zanardi2002,Dowling2006}. This has led to a detailed discussion on the question of the role of exchange symmetry in generating entanglement in these systems \cite{ECKERT200288,Wiseman2003,PhysRevA.70.012109,PhysRevLett.110.080503,PhysRevA.67.024301,PhysRevA.64.054302,PhysRevLett.112.150501} and whether it can be exploited for quantum information tasks \cite{franco2016quantum,PhysRevLett.120.240403,morris2019entanglement}. This debate has gained new urgency through the advent of advanced experiments that operate on individually controllable, indistinguishable particles such as photons or ultracold atoms and can record single-particle-resolved correlations. 
Distinguishing correlations rooted in interactions from those originating from quantum statistics and quantifying entanglement are key challenges in these experiments \cite{PhysRevLett.93.110501,PhysRevLett.109.020505, Cramer,Islam,Preiss1229,bergschneider2018correlations,Jeltes2007,Bocquillon1054,Fadel409,Lange416,Kunkel413}.

\begin{figure}
	\centering
	\includegraphics{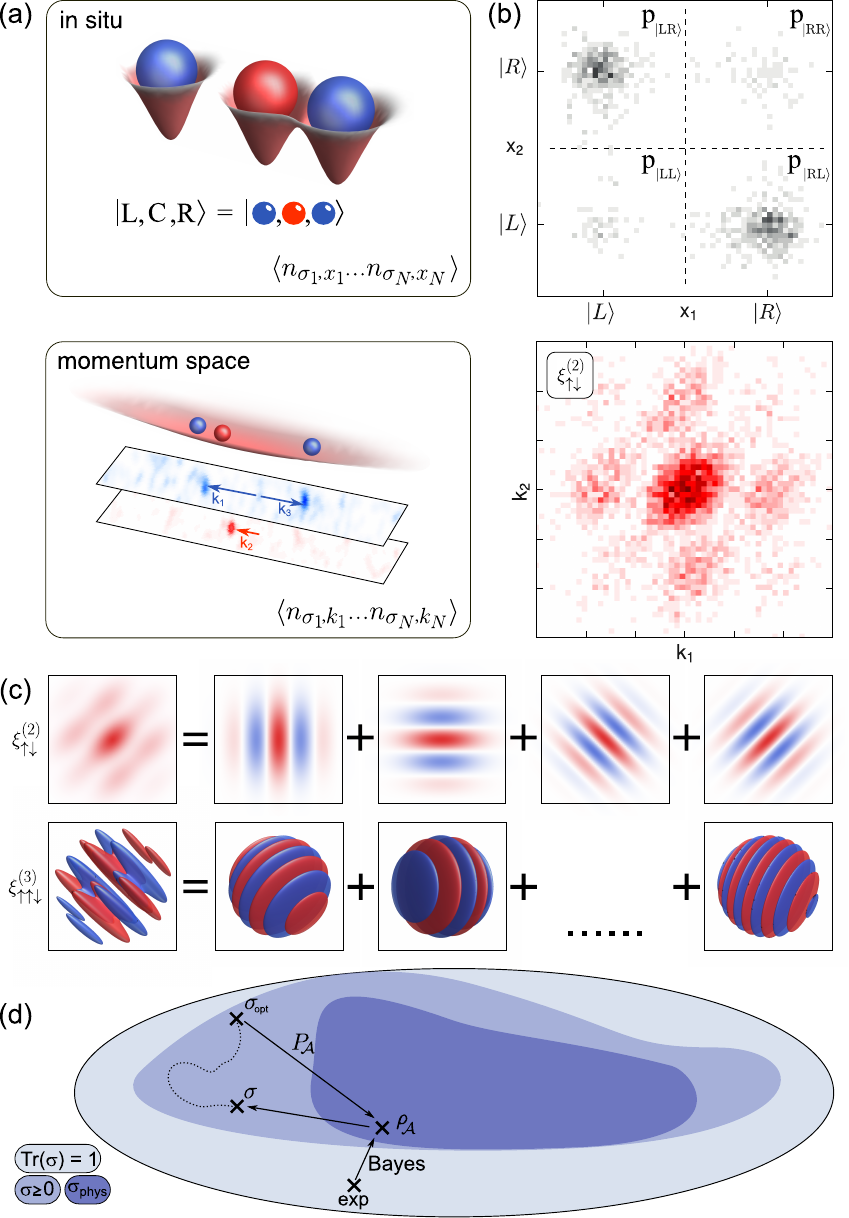}
	\caption{State reconstruction scheme. (a) We initialize atoms in an array of optical tweezers and measure single-atom and spin resolved in situ populations and momentum space correlation functions. (b) Example for in situ and momentum correlation functions for two atoms in two tweezers. (c) We decompose momentum correlation functions into a set of trigonometric basis functions, as shown for $N=2$ (upper row) and $N=3$ (lower row). The weight of the basis functions is used to constrain off-diagonal entries of the density matrix.	(d) The bare experimentally measured density matrix may lie outside the space of physical density matrices. We use Bayesian state estimation to construct a density matrix $\rho_A$ that is both positive semidefinite and obeys exchange antisymmetry. The $\AN$ searches the larger space of all physical density matrices, irrespective of their symmetries, for the smallest negativity consistent with $\rho_A$.}
	\label{fig1}
\end{figure}
If one is interested in entanglement between particles rather than spatial modes, useful approaches can be formulated in first-quantised notation. Labels $1\ldots N$ are introduced to identify particles, which makes it possible to treat them as subsystems and apply well-known entanglement measures. However, great care must be taken regarding the meaning of particle labels, given the impossibility of physically differentiating between indistinguishable particles. For instance, let us consider two spinless fermions in two spatially separated modes, labeled $L$ and $R$. Due to the indistinguishability of the two fermions, the first-quantised wavefunction has to be antisymmetrised over the two spatial modes, $\ket{\Psi_\text{as}} = \frac{1}{\sqrt{2}}\left( \ket{L}_1\ket{R}_2 - \ket{R}_1\ket{L}_2\right)$, where $\ket{\cdot}_i$ describes the spatial mode of the $i^{th}$ particle. Formally this state resembles a maximally entangled Bell state. However, in the first-quantised formalism the antisymmetrisation simply represents the observer's inability to distinghuish the two particles. The entanglement can hence be argued to be an artefact of attributing artificial labels to the atoms \cite{GMW}. 
Nevertheless the antisymmetrization requirement of the wavefunction can induce strong correlations in experimental observables and it is desirable to analyze experimental measurements in such a way that symmetrisation effects can be separated from other forms of entanglement. This calls for new ways of evaluating entanglement from experimental data, particularly in the presence of imperfections and noise. \\
Here, through measurements made on multi-well few-atom systems, we explore the relationship between fermionic exchange antisymmetry and indistinguishable particle entanglement. To this end, we introduce the notion of an antisymmetric negativity, in which we determine the minimum entanglement that must already exist in a state before symmetrisation in order to describe the experimental quantum state. This quantitatively separates the notions of entanglement due to interactions and correlations due to symmetrisation.

According to the criteria laid out by Ghirardi, Marinatto, and Weber \cite{GMW}, an identical-particle pure antisymmetric state $\ket{\psi_\mathcal{A}}$ that can be obtained by antisymmetrising a product state is separable and hence should not be considered to be entangled. Building on this intuition we propose to quantify entanglement of a fermionic density matrix $\rho_\cA$ through the functional
\begin{equation}
    \mathcal{E}_\cA(\rho_\cA)=\min\limits_\sigma\left\lbrace\mathcal{E}(\sigma): P_\cA \sigma P_\cA = \frac{1}{2} \rho_\cA \right\rbrace,
    \label{AN}
\end{equation}
with the projector $P_\cA$ on the antisymmetric subspace and any entanglement measure $\mathcal{E}(\sigma)$. As illustrated in Fig.\,\ref{fig1}\,(d), the optimization variable $\sigma$ is a normalized quantum state that does not obey any specific exchange symmetry and exists in a larger Hilbert space than the antisymmetric (physical) state $\rho_\cA$. If there exists a product state $\sigma$ whose projection on the antisymmetric subspace is $\rho_\cA$, then $ \mathcal{E}_\cA(\rho_\cA) = 0$. Otherwise, if there is no such state, then $ \mathcal{E}_\cA(\rho_\cA) > 0$ and the identical particle state ought to be regarded as entangled. For the case of $\mathcal{E}$ being the standard negativity $\mathcal{N}$ \cite{PhysRevA.65.032314}, Eq.(\ref{AN}) can be explicitly calculated in the form of a semidefinite program \cite{enrico1,enricoPhD,Methods} and is called the \textit{Antisymmetric Negativity} $\mathcal{N}_\cA(\rho_\cA)$ ($\AN$). For instance if we consider the state of two identical fermions, we obtain $\ket{\Psi_\text{as}}$ by antisymmetrizing the product state $\ket{L}_1\ket{R}_2$, which has a negativity of zero and therefore also the $\AN$ is zero. The $\AN$ hence allows us to treat particles as subsystems of a many-body state without ascribing entanglement to correlations arising from antisymmetrisation alone.\\
To benchmark the usefulness of the $\AN$, we develop a general scheme to obtain physical density matrices of few-atom systems from experimental data and to calculate lower bounds on the $\AN$. This experimental technique has in parts already been presented in previous publications \cite{PhysRevA.97.063613,bergschneider2018correlations,highcontrast}. The key features we demonstrate are: (i) The $\AN$ can be computed from experimental data including noise; (ii) it identifies states which exhibit correlations only due to quantum statistics as unentangled; (iii) it identifies interaction-driven entanglement in two-particle systems, and (iv) it can be extended to multipartite scenarios.
\begin{figure}
	\centering
	\includegraphics{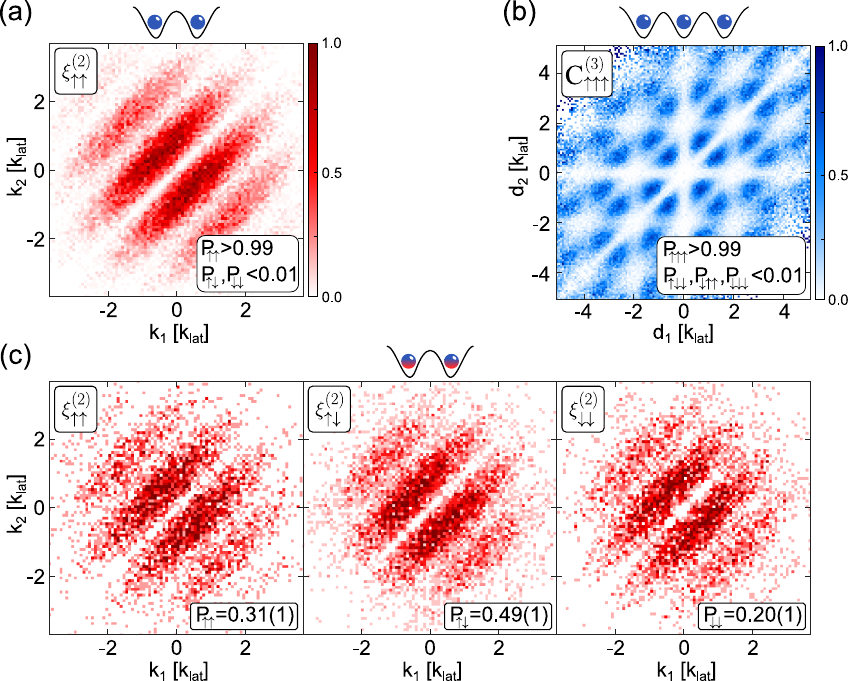}
	\caption{Momentum correlation measurements of states (a) $\ket{L\uparrow}\ket{R\uparrow}$, (b) $\ket{L\uparrow}\ket{C\uparrow}\ket{R\uparrow}$, and (c) $\ket{L\rightarrow}\ket{R\rightarrow}$. The latter is created by rotating $\ket{L\uparrow}\ket{R\uparrow}$ with a radio-frequency pulse, which results in strong second-order momentum correlations in all spin combinations. We calculate $\mathcal{N}_2(\rho_a) = 0.5$, $\mathcal{N}_2(\rho_b) = 1$, $\mathcal{N}_2(\rho_c) = 0.60^{+0.13}_{-0.11}$ and $\AN_2(\rho_a) = \AN_2(\rho_b) = 0$, $\AN_2(\rho_c) = 0.048^{+0.064}_{-0.048}$ for the respective states. The $\AN$ identifies all states as separable.}
	\label{fig2}
\end{figure}

We use the $\AN$ for a quantum state characterization in few-body Fermi-Hubbard systems. Experimentally, the states of interest are realized with interacting $^6$Li atoms in two different hyperfine states (labeled $\ket{\uparrow}$ and $\ket{\downarrow}$) trapped in an array of optical tweezers, see Fig.\,\ref{fig1}(a). We probe the states via correlation measurements and start the investigation of the system by measuring single-particle and spin resolved real-space and momentum-space distributions. Within the in situ dataset we analyze the appearance of all possible spin combinations and restrict the further analysis to the relevant spin sector of the full Hilbert space. From the obtained in situ measurements we also extract the diagonal entries of the density matrix (Fig.\,\ref{fig1} (b)). In the next step we calculate momentum-space correlation functions of $N^{th}$ order $\xi^{(N)}_{\sigma_1 ... \sigma_N} = \langle n_{\sigma_1,k_1}... n_{\sigma_N,k_N}\rangle$ (Fig.\,\ref{fig1}(a) \& (b)), which we decompose into a set of trigonometric basis functions \cite{Methods}, see Fig.\,\ref{fig1}(c). Following the scheme from \cite{bergschneider2018correlations,Bonneau}, we relate the weight of the basis functions to off-diagonal entries of the density matrix \cite{Methods}. Since this does not constrain the full density matrix and in order to prevent unphysical properties due to statistical and systematic measurement uncertainties, we perform a Bayesian quantum state estimation \cite{BME} based on the constrains that are obtained from the experimental data. We convert the experimental density matrices into their first quantised representation \cite{Methods} and use the full posterior distribution of density matrices for the calculation of the $\AN$. All values of the $\AN$ are presented as the median together with a $68\%$ credible interval. 

\begin{figure}
    \centering
    \includegraphics{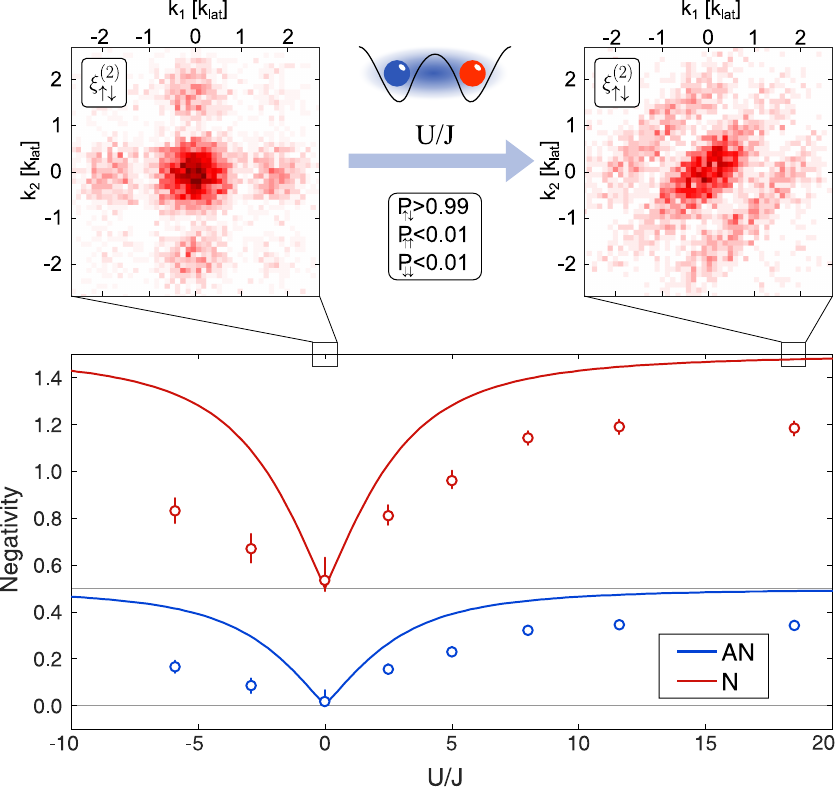}
    \caption{Entanglement as a function of interaction strength in the ground state. The plot shows both the standard negativity and the antisymmetric negativity as a function of interaction strength $U/J$ calculated from experimentally obtained density matrices (circles) and the theoretical expectations (solid lines) for the ground state of a double well potential. Technical noise leads to a loss of purity and a reduction in experimentally detected negativity with respect to the pure state expectation. }
    \label{fig3}
\end{figure}

Figure \ref{fig2} shows measurements of momentum correlations functions for three states that were obtained by independently preparing a single fermion in each optical tweezer. For the three-particle state, we show the correlation function after integration, i.e. 
\begin{equation}
    C_{\sigma_1\sigma_2\sigma_3}(d_1,d_2) = \frac{\int{\langle n_{\sigma_1, k} n_{\sigma_2, k+d_1} n_{\sigma_3, k+d_2}\rangle dk}}{\int{\langle n_{\sigma_1, k}\rangle \langle n_{\sigma_2, k+d_1}\rangle \langle n_{\sigma_3, k+d_2}\rangle dk}}.
\end{equation}

\noindent The states are spin polarized along the $z$-axis ($\ket{L\uparrow}\ket{R\uparrow}$ (a), $\ket{L\uparrow}\ket{C\uparrow}\ket{R\uparrow}$ (b)), or the $x$-axis ($\ket{L\rightarrow}\ket{R\rightarrow}$ (c)). Our measurement is performed in the $\sigma_z$ basis, so that the $x$-polarized state shows spin fluctuations, while the two other states do not. All states show strong momentum correlations, in particular, the $x$-polarized state shows correlations in all three spin combinations.
We apply the reconstruction scheme and calculate negativities for all states. 
The calculation of the standard negativity from the first-quantised density matrix yields a value larger than zero because it interprets the antisymmetrised structure of the first-quantized state as entanglement between individual particles. In contrast, the $\AN$ returns values consistent with zero (see caption of Fig.\,\ref{fig2}). This demonstrates that the antisymmetric negativity identifies those states as separable and hence unentangled, while the standard negativity detects particle-particle entanglement also when it is due to antisymmetrisation.

To test the $\AN$ on states where correlations are substantially modified by interaction, we investigate two atoms of different spin in the ground state of a double-well potential with tunnel coupling $J$ and tunable on-site interaction $U/J$ \cite{bergschneider2018correlations}. The two atoms are prepared in a spin-singlet configuration, $\ket{S,m_S} = \ket{0,0}$, so that correlations appear only between opposite spins. Figure \ref{fig3} summarizes the results of the measurements together with the theoretically expected negativities for pure states. Both negativities are smallest in the non-interacting case and increase with both attractive and repulsive interactions. $\mathcal{N}$ is always larger than $\AN$ and even non-zero for vanishing interactions. The finite value of $\mathcal{N}$ at $U/J = 0$ is caused by the antisymmetrisation of the spin degree of freedom. In contrast the $\AN$ identifies the state as separable, since the wave function can be obtained by antisymmetrising the product of two non-interacting atoms in the ground state of the double-well.
In this sense, the $\AN$ identifies entanglement that is induced by interactions. We note that the $\AN$ takes exactly the same values as the spin-mode negativity \cite{bergschneider2018correlations}. In this case the particles can be uniquely identified by their spin state.

\begin{figure}
    \centering
    \includegraphics{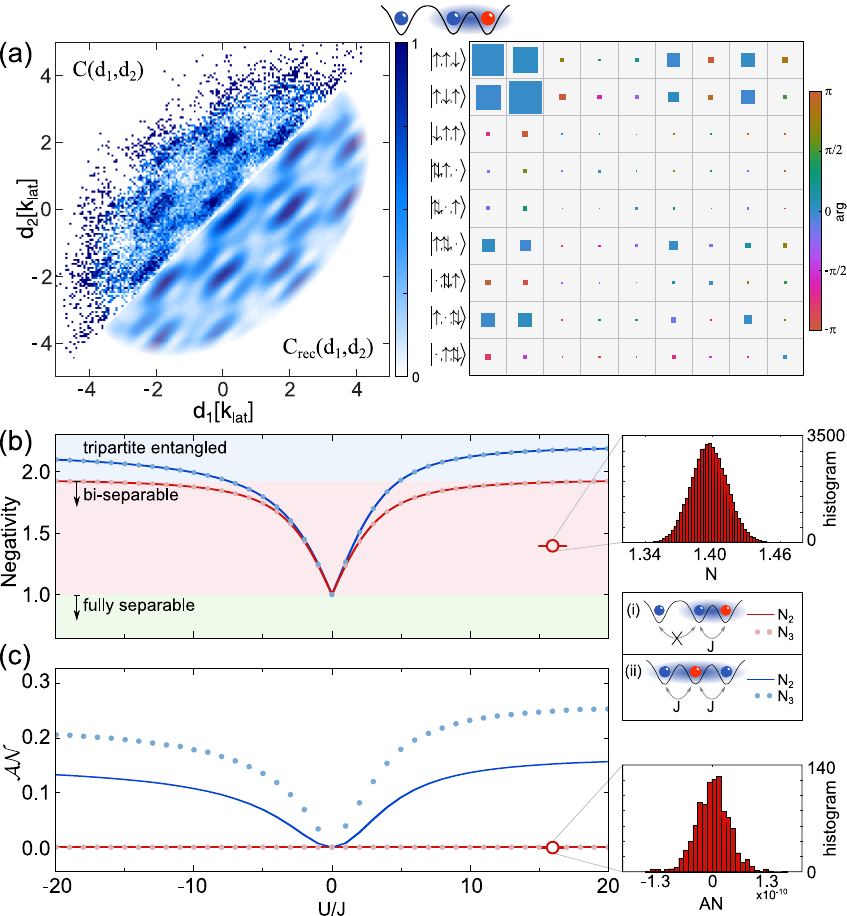}
    \caption{Multipartite generalization of $\AN$. (a) Measured and reconstructed $C_{\downarrow\uparrow\uparrow}(d_1,d_2)$ and reconstructed density matrix $\rho_\text{BME}$ as Hinton diagram for a three-particle state with an interacting singlet in a double well and an independent particle in the leftmost well. (b) Bi- and tripartite negativities $\mathcal{N}_2/\mathcal{N}_3$ for shown states (see legend) for pure theoretical density matrices and experimentally reconstructed states. Coloured regions indicate negativity witnesses as discussed in the main text. (c) Bi- and tripartite $\AN_2/\AN_3$ for the same states as in (b). The histograms show the distribution of $\mathcal{N}_2$ and $\AN_2$ over the posterior distribution from the Bayesian state estimation.}
    \label{fig4}
\end{figure}

We now generalize the scheme to the case of three atoms, where strong correlations may arise both due to antisymmetrisation and interaction at the same time. We experimentally initialize a three-atom state in a triple-well potential with $J_{LC} = 0$ and $J_{CR}=J$, see inset (i) of Figure \ref{fig4}\,(b). 
The left well is entirely decoupled from the other wells and initialized with a single particle, which is correlated with the interacting double well only through antisymmetrization. \\
Figure \ref{fig4}\,(a) shows the measured and the reconstructed momentum correlator $C_{\downarrow\uparrow\uparrow}(d_1,d_2)$ together with the physical density matrix $\rho_\text{BME}$ that we obtain from the Bayesian state estimation, where $\rho_\text{BME}$ is the mean of the posterior distribution. The state exhibits complex momentum correlations that are caused both by antisymmetrisation and interaction. 
In addition to the experimentally realised state at $U/J=16$ we theoretically investigate the same state for various interaction strengths from strongly attractive ($U/J=-20$) to strongly repulsive ($U/J=20$) interactions. We also investigate theoretically the ground state of the triple well with homogeneous couplings where all three particles tunnel and interact (as illustrated in the legend (ii) in Fig.\,\ref{fig4} (b-c)). The two lower panels in Fig.\,\ref{fig4} summarize the calculated bipartite and tripartite \cite{Methods} $\mathcal{N}_{2}$/$\mathcal{N}_{3}$ (b) and $\AN_{2}$/$\AN_{3}$ (c) for the experimental and the theoretical states. We find identical values for both bi- and tripartite negativities due to the antisymmetrized structure of the density matrices.

A first characterization of the theoretically expected entanglement of both states can be made by using the standard negativity $\mathcal{N}$ as a witness for bi- and tripartite entanglement. It can be shown that for three particles, the maximum negativity that can be obtained from antisymmetrising a fully separable state is 1, whereas an antisymmetrised biseparable state exhibits a maximum negativity of 1.94 \cite{Methods}. Indeed the first state (i) saturates these limits for $U/J=0$ and $U/J\rightarrow \pm\infty$, respectively. For the experimental state we find $\mathcal{N} = 1.39\pm0.02$, exceeding the bound for fully separable states. The second state (ii), on the other hand, exhibits a theoretical negativity that for large attractive and repulsive interactions exceeds 1.94. This is a sufficient condition to certify genuine multipartite entanglement. 

A complete picture of the presence of multipartite entanglement, however, is only provided by the $\AN$: For state (i), the minimization finds a state where a single atom can be factorized out, e.g. for the partitioning $\mathcal{N}_{12|3}$ and $U/J\rightarrow \infty$ the optimal state $\sigma_\text{opt} = \left(\ket{C\uparrow}_1 \ket{R\downarrow}_2 + \ket{R\downarrow}_1 \ket{C\uparrow}_2\right)\ket{L\uparrow}_3$ is found. At $U=0$, the search returns a fully separable three-particle state. Hence both $\AN_2$ and $\AN_3$ evaluate to zero for all interaction strengths, indicating the overall absence of genuine multipartite entanglement due to the factorizable nature of the state. Indeed for any biseparable state (of arbitrary particle number), the $\AN$ is zero, if the size of the subsystems coincides with the size of the partitioning under which the state is biseparable. For state (ii) the $\AN_2$ and $\AN_3$ are positive for any interaction strength indicating the presence of both bi- and tripartite entanglement even at small but non-zero interaction. This conclusion is not possible from the standard negativity alone. 
To assess the relevance of he $\AN$ for experimental investigations of tripartite entanglement we finally analyze its robustness against white noise. We start with the ground state of the triple well for $U/J = 16$ and gradually add white noise, $\rho_{noise} = (1-r) \mathbb{1} + r\ket{\Psi}\bra{\Psi}$, with $r\in[0,1]$. We find that tripartite entanglement persists down to $r=0.75$, which corresponds to a purity of $p = 0.61$. Such purities can reasonably be expected to be achieved in experiments, confirming the applicability of multipartite AN to relevant experimental scenarios. 

In this work we introduced a new notion of entanglement for fermionic density matrices, the antisymmetric negativity. Our approach explicitly quantifies entanglement between particles, rather than spatial modes and distinguishes correlations from antisymmetrization from those induced by interactions. We demonstrated a scheme to extract physical density matrices from experimental data and apply the $\AN$ to experimentally reconstructed density matrices of a few mobile atoms in an array of optical tweezers. The $\AN$ correctly identifies separable states as unentangled while the standard negativity detects the antisymmetrized structure of the density matrix as entanglement. The $\AN$ can also detect genuine tripartite entanglement and can be used to reveal the separability structure of an antisymmetrized fermionic density matrix. Our results enable the quantitative study of entanglement and exchange antisymmetry in experimental settings including quantum dot arrays \cite{Hensgens2017}, atomic or molecular systems, and cold atom experiments, and helps to provide further structure and insight into the entanglement of indistinguishable particles in these experimental settings.

\begin{acknowledgments}
We would like to thank Marco Piani for his contribution to the early stages of this project, particularly in relation to the notion of antisymmetric negativity. This work was supported by the European Union’s Horizon 2020 research and innovation program under  grant agreement No.~817482 PASQuanS. Work at the University of Strathclyde was supported by the EPSRC Programme Grant DesOEQ (EP/P009565/1). The experimental work has been supported by the ERC consolidator grant 725636, the Heidelberg Center for Quantum Dynamics, and the DFG Collaborative Research Centre SFB 1225 (ISOQUANT). PMP acknowledges funding from the Daimler and Benz Foundation. 
\end{acknowledgments}

\clearpage

\section{METHODS}
	
	\subsection{Experimental Setup}
	Experiments are conducted with two of the three lowest hyperfine states in $^6$Li, that we trap in a multiwell potential. This potential is created from individually controllable optical tweezers at a wavelength $\lambda = 1064 \si{n\meter}$, with a waist of $\omega = 1.15\si{\mu \meter}$, and tunable distance and depth. We start by preparing a deterministic number of atoms in our tweezer traps \cite{serwane,DWSimon} and tune the interaction between different spin states via a broad $s$-wave Feshbach resonance to eventually induce entanglement between different spin states \cite{bergschneider2018correlations}. After initializing a state of interest with $N$ atoms, we measure spin-resolved real-space populations and momentum-space correlation functions of $N^{th}$ order $\xi^{(N)} = \langle n_{\sigma_1}(k_1)... n_{\sigma_N}(k_N)\rangle$, by repeatedly measuring spin-resolved single-particle momenta by applying resonant light and collecting the fluorescence signal on a single-photon sensitive EMCCD camera \cite{PhysRevA.97.063613}. We postselect the images to the correct atom number with postselection rates of $\approx$ 80\% (70\%) for states with 2(3) atoms. In-depth experimental details can be found in our previous publications \cite{highcontrast,bergschneider2018correlations,PhysRevA.97.063613}.
	
	\subsection{Density Matrix Reconstruction}
	
	To construct density matrices from our experimental data we follow the procedure described in our previous work \cite{bergschneider2018correlations}: Diagonal entries $\rho_{ii}$ are directly extracted from the measured in situ populations.
	In order to access the off-diagonal entries of the density matrix, we decompose the measured correlation function $\xi^{(N)}$ into a set of trigonometric basis functions $\{\mathcal{\mathbf{C}}, \mathcal{\mathbf{S}}\}$
	\begin{align}
		\xi^{(N)} &= \sum_i \left( c_i \mathcal{C}_i + s_i \mathcal{S}_i \right )\nonumber\\ 
		&= \sum_i \left(c_i \cos{(2 \pi \vec{q}_i\vec{k})}+s_i \sin{(2 \pi \vec{q}_i\vec{k})}\right ).
		\label{decomposition}
	\end{align}
	Here $\vec{k} = (k_1,k_2,...,k_N)^T$ and $\vec{q}_i$ is a vector with real-space distances between spatial modes.
	
	We also analytically calculate the $N^{th}$ order momentum correlation functions for an arbitrary density matrix $\rho$ \cite{Bonneau,bergschneider2018correlations}. The real part of the density matrix contributes to the cosine patterns $\mathcal{C}_i$, whereas the imaginary part contributes sine patterns $\mathcal{S}_i$. The full momentum correlation function can be written as 
	
	\begin{equation}
	\xi^{(N)} = \sum_{i,j} M_{ij} \Re{(\rho_j)} \mathcal{C}_i + N_{ij} \Im{(\rho}_j) \mathcal{S}_i.
	\end{equation}
	
	\noindent Here, $\mathbf{M}$ and $\mathbf{N}$ are coefficient matrices and $\mathbf{\rho}$ is a vectorized form of the density matrix. 
	
	Equating this form of the momentum correlator to the experimentally obtained coefficients from Eq.(\ref{decomposition}), one obtains a set of linear equations
	\begin{align}
		\mathbf{M}\cdot\Re{(\mathbf{\rho})} = \mathbf{c} \nonumber \\
		\mathbf{N}\cdot\Im{(\mathbf{\rho})} = \mathbf{s} 
		\label{eqn:linear_combinations}
	\end{align}

	This scheme does not fully constrain the density matrix \cite{bergschneider2018correlations}. In combination with in situ measurements, the system of equations (\ref{eqn:linear_combinations}) constrains 12 of 16 parameters in the two-particle density matrix and 69 of 81 parameters in the three-particle density matrix. In order to find the physical density matrix that is most compatible with our measurements, we perform a Bayesian quantum state estimation \cite{BME}, taking the full set of equations (\ref{eqn:linear_combinations}) into account. 
	The Bayesian state estimation returns a sample set of physical density matrices whose distribution represents our full knowledge from real and momentum space measurements. The spread of density matrices within this set represents both the experimental error in the measured coefficients \textbf{c}, \textbf{s} as well as the uncertainty associated with measuring an incomplete set of parameters of $\rho$. An example of a reconstructed density matrix is shown in Fig.\,4 (c) of the main text. We verify the validity of $\rho_{BME}$ by computing the momentum correlations such a state would have and comparing them to the measured $\xi^{(N)}$ and find good agreement.

	\subsection{Data Preparation}
	In order to address identical-particle entanglement we need to write the experimentally obtained density matrices $\rho_{exp}$ in the first-quantized representation. For the double-well data this is achieved by mapping $\rho_{exp}$ onto the 16-dimensional basis $\{ \ket{X \sigma}\ket{Y \tau}\}$, for $X,Y=L,R$ and $\sigma,\tau=\uparrow,\downarrow$. The triple well is mapped onto the 216-dimensional basis $\{ \ket{X \sigma}\ket{Y \tau} \ket{Z \kappa}\}$, for $X,Y,Z=L,C,R$ and $\sigma,\tau, \kappa=\uparrow,\downarrow$. 
	
	The substitution rule is given by the slater determinant, i.e. $\ket{X}_\uparrow\ket{Y}_\downarrow \rightarrow \frac{1}{\sqrt{2}}(\ket{X\uparrow}\ket{Y\downarrow}-\ket{Y\downarrow}\ket{X\uparrow})$ for the double-well case and analogously for the triple-well case.

	\subsection{Antisymmetric Negativity}
	According to the GMW criterion \cite{GMW}, identical fermions of a composite quantum system that are described by a pure state $\ket{\Psi_\mathcal{A}}$ are non-entangled iff the wave function can be obtained by antisymmetrizing a factorized state $\Psi_{sep}$,
	\begin{equation}
	\frac{P_\mathcal{A} \ket{\Psi_{sep}}}{||P_\mathcal{A} \ket{\Psi_{sep}}||} = \ket{\Psi_\mathcal{A}},
	\end{equation}
	with $P_\mathcal{A}$ the projection operator onto the antisymmetric subspace. A general way of quantifying entanglement between two subsystems $A$ and $B$ is given by the negativity. For a first quantized state $\rho$, the negativity is given by
	\begin{equation}
	\mathcal{N} = \frac{||\rho^{\Gamma_A}|| - 1}{2}
	\end{equation}
	where $\rho^{\Gamma_A}$ is the partial transpose of $\rho$ with respect to subsystem $A$, and $||\cdot||$ denotes the trace norm. When treating particles (labelled 1..N) as subsystems, the standard negativity of the above form leads to conceptual difficulties.
	To remedy these problems and to investigate entanglement properties of a fermionic identical-particle state, we introduce a new concept that reverses the antisymmetrization before quantifying entanglement \cite{enricoPhD}. \\
	Starting from the functional from Eq.(1) of the main text we define the \textit{Antisymmetric Negativity} (AN) as
	\begin{equation}
	\mathcal{N}_\cA(\rho_\cA)=\min\limits_\sigma\left\lbrace\mathcal{N}(\sigma):\, P_\cA \sigma P_\cA = c \rho_\cA \right\rbrace,
	\label{ANSOM}
	\end{equation}
	where $c=\max\{\ttr(P_\cA \sigma):\,\sigma=\sigma_\mathrm{PPTm} \}$ is the maximal projection probability of $\sigma$ such that $\rho_\cA$ may not be obtained by antisymmetrizing a PPT-mixed state. As derived in \cite{enricoPhD,PhysRevA.63.042111} we use $c=\frac{1}{2}$ for the double-well states and $c= \frac{1}{3}$ for the triple-well states.\\
	The AN of a fermionic density matrix $\rho_\cA$ is the smallest negativity $\mathcal{N}$ possible for an unsymmetrized density matrix $\sigma$ that generates $\rho_\cA$ under antisymmetrization. Besides being positive semidefinite, $\sigma \geq 0$, there is no further constraint on $\sigma$ such that also states without a particular exchange symmetry are taken into account. 
	This extensive search can be formulated as a \textit{semidefinite program} (SDP) \cite{enrico1,enricoPhD} and explicitly calculated.
	
	Semidefinite programming finds numerous applications in the theory of quantum computing \cite{watrous_2018} since it can be efficiently solved by interior point methods \cite{boyd2004convex}. A SDP, consisting of a triple $(A,B,\Phi)$, is a convex optimization algorithm minimizing a linear function $\langle A,X\rangle$ with hermicity-preserving constraints $B = \Phi(X)$ on hermitian matrices X. It is commonly written in the form
	\begin{align}
		\text{minimize  } & &\langle A,X \rangle \nonumber \\
		\text{such that  } && \Phi(X) = B \nonumber \\
		&& X\geq 0.
	\end{align}
	
	\noindent Rewriting the trace norm as a SDP \cite{watrous_2018,enricoPhD} suggests that also the bipartite negativity $\mathcal{N}_2$ can be written in this form,
	
	\begin{align}\label{sdp:neg}
		\text{minimize} &&(\ttr(M)-1)/2  \nonumber \\
		\text{such that} && -M\leq \rho^\Gamma \leq M. \nonumber \\
		&& M \geq 0
	\end{align}
	
	\noindent Here, $\rho^\Gamma$ is the partial transpose of the density matrix and the negativity is the optimal value of the minimization process. In such an optimization, measurement outcomes may be enforced as additional constraints such that the optimal value will be the minimum negativity compatible with measurement outcomes. In the same way as for the negativity we can express the bipartite antisymmetric negativity $\mathcal{AN}_2$ as a SDP and explicitly calculate it for any bipartite state. For this we extend Eq.(\ref{sdp:neg}) to
	\begin{align}\label{sdp:ipe}
		\text{minimize} && (\ttr(M)-1)/2  \nonumber \\
		\text{such that}
		& &  -M\leq X^\Gamma \leq M  \nonumber\\
		& & M \geq 0\nonumber \\
		& & X\geq 0  \nonumber \\
		& & \ttr(X) = 1  \nonumber \\
		& & P_\cA X P_\cA = \ttr (P_\cA X) \rho_\cA  \nonumber \\
		& & \ttr (P_\cA X)=c.
	\end{align}
	\noindent This SDP searches for the smallest negativity of a hermitian matrix $X \geq 0$ with $\ttr{(X)} = 1$, whose projection on the fermionic density matrix is $c$. 
	
	Generalizing entanglement negativity to genuine multipartite entanglement \cite{Jungnitsch2011,Hofmann_2014} also enabels the generalization of the AN beyond the bipartite case. The idea is to minimize the bipartite negativity over all bipartitions of the quantum system, as well as for all possible convex combinations of density matrices. For our triple well case, given an (anti)symmetric state $\rho_\cA$, we call genuine tripartite negativity $\cN_3(\rho_\cA)$ the optimal value for the SDP
	\begin{align}
		\text{minimize}&&(\mathrm{Tr}(M_{A|BC}+M_{B|AC}+M_{C|AB})-1)/2 \nonumber \\
		\text{such that}
		&& M_{A|BC},M_{B|AC},M_{C|AB}\geq 0\nonumber \\
		&& \tilde{\rho}_{A|BC},\tilde{\rho}_{B|AC},\tilde{\rho}_{C|AB}\geq 0\nonumber \\
		&& -M_{A|BC}\leq \tilde{\rho}_{A|BC}^\Gamma \leq M_{A|BC} \nonumber \\
		&& -M_{B|AC}\leq \tilde{\rho}_{B|AC}^\Gamma \leq M_{B|AC} \nonumber \\
		&& -M_{C|AB}\leq \tilde{\rho}_{C|AB}^\Gamma \leq M_{C|AB} \nonumber \\
		&& \rho_\cA=\tilde{\rho}_{A|BC}+\tilde{\rho}_{B|AC}+\tilde{\rho}_{C|AB},  
	\end{align}
	and genuine tripartite antisymmetric negativity $\mathcal{AN}_3(\rho_\cA)$ the optimal value for the SDP,
	\begin{align}\label{sdp:gmnipe}
		\text{minimize}&&(\mathrm{Tr}(M_{A|BC}+M_{B|AC}+M_{C|AB})-1)/2 \nonumber \\
		\text{such that}
		&& M_{A|BC},M_{B|AC},M_{C|AB}\geq 0\nonumber \\
		&& \tilde{\rho}_{A|BC},\tilde{\rho}_{B|AC},\tilde{\rho}_{C|AB}\geq 0\nonumber \\
		&& -M_{A|BC}\leq \tilde{\rho}_{A|BC}^\Gamma \leq M_{A|BC} \nonumber \\
		&& -M_{B|AC}\leq \tilde{\rho}_{B|AC}^\Gamma \leq M_{B|AC} \nonumber \\
		&& -M_{C|AB}\leq \tilde{\rho}_{C|AB}^\Gamma \leq M_{C|AB} \nonumber \\
		&& \tilde{\rho}=\tilde{\rho}_{A|BC}+\tilde{\rho}_{B|AC}+\tilde{\rho}_{C|AB}  \nonumber \\
		&& P_\cA \tilde{\rho} P_\cA = \ttr (P_\cA \tilde{\rho}) \rho_\cA \nonumber \\
		&& \ttr (P_\cA \tilde{\rho}) = c,
	\end{align}
	
	\noindent where $c=1/3$ for the triple-well case \cite{PhysRevA.63.042111}. The indices indicate the partitioning of the system in the space of density matrices before antisymmetrization. $A|BC$ means that the partial transposition operation is carried out on subsystem A for the partition $A|BC$. Equation (\ref{sdp:gmnipe}) can be extended to the general N-partite case.

	We solve all problems using Qetlab \cite{qetlab} and CVX, a package for specifying and solving convex programs \cite{cvx,gb08}.
	We compute $\mathcal{N}_2/\mathcal{N}_3$ and $\mathcal{AN}_2/\mathcal{AN}_3$ over a representative subset of all states returned by the state estimation and give the median and the 68\% credible interval of the obtained distribution in the main text.

	\subsection{Negativity Bounds}
	
	A fully separable pure three-atom state can be written as
	\begin{equation}\label{eq:sep}
	\ket{\psi_\mathrm{sep}^p}=\ket{\psi}\otimes\ket{\phi}\otimes\ket{\chi},
	\end{equation}
	with $\ket{\psi}$, $\ket{\phi}$, and $\ket{\chi}$ being pure single-particle states.
	After antisymmetrisation of $\psi_\mathrm{sep}^p$, the normalized state reads as
	\begin{equation}
	\rho^{\mathcal{A}}_\mathrm{sep}=\frac{
		P_\mathcal{A}\ket{\psi^p_\mathrm{sep}}\bra{\psi^p_\mathrm{sep}}P_\mathcal{A}}{\ttr{\left(P_\mathcal{A}\ket{\psi^p_\mathrm{sep}}\bra{\psi^p_\mathrm{sep}}P_\mathcal{A}\right)}},
	\end{equation}
	with a bipartite negativity of $\mathcal{N}_2(\rho^\mathcal{A}_\mathrm{sep}) = 1$ along any possible bipartition due to the exchange symmetry.
	A general mixed state which is obtained by antisymmetrizing a separable state may be written as
	\begin{equation}\label{eq:sepa}
	\tilde{\rho}^\mathcal{A}_\mathrm{sep}=\sum_i p_i\rho^{\mathcal{A}}_\mathrm{sep,i}.
	\end{equation}
	Due to the convexity of entanglement negativity, we have the inequality
	\begin{equation}\label{eq:ineq}
	\mathcal{N}(\tilde{\rho}_\mathrm{sep}^\cA) \leq \sum_i p_i  \mathcal{N}_2\left(\rho^{\mathcal{A}}_\mathrm{sep,i} \right)=1,
	\end{equation}
	
	Therefore, we have the upper bound for the bipartite negativity of an antisymmetrised fully separable state,
	\begin{equation}
	\mathcal{N}^{\max}_\mathrm{sep}=1.
	\end{equation}
	
	In a similar fashion, we may address the maximum negativity of the antisymmetrization of a biseparable state. A pure biseparable state can be written as
	
	\begin{equation}\label{bisep}
	\ket{\psi_\mathrm{bs}^p}=\ket{\psi}_1\otimes\ket{\phi}_{23},
	\end{equation}
	where $\ket{\phi}_{23}$ is any pure two-particle state between atoms with labels 2 and 3.
	We can write a maximally entangled state of the form
	\begin{equation}\label{eq:bs}
	\ket{\phi}_{23}=\frac{1}{\sqrt{2}}\left( \ket{X\sigma}\ket{Y\tau}+e^{i\theta}\ket{Y\tau}\ket{X\sigma} \right),
	\end{equation}
	with $X,Y \in \{L,C,R\}$ and $\sigma,\tau \in \{\uparrow,\downarrow\}$. We explicitly calculate the negativity of its antisymmetrised form  $\rho_\mathrm{bs}^\cA$ to be $\mathcal{N}_{2,\mathrm{bs}}\equiv \mathcal{N}_2(\rho_\mathrm{bs}^\cA)\simeq 1.9428$.
	Based on the argument underlying equation \eqref{eq:ineq}, any mixed state which is obtained by antisymmetrizing a convex combination of biseparable states of the form \eqref{eq:bs} will have a bipartite negativity upper bounded by $\mathcal{N}_{2,\mathrm{bs}}$.

\end{document}